\documentclass[twocolumn,superscriptaddress]{revtex4-1}

\usepackage{array}
\usepackage{amssymb}
\usepackage{amsmath}
\usepackage{graphicx}

\begin{document}

\title{Viscoelastic transient of confined Red Blood Cells}

\author{Ga{\"e}l Prado} 
\affiliation{LIPhy, CNRS, F-38000, Grenoble, France}
\affiliation{LIPhy, Univ. Grenoble Alpes, F-38000, Grenoble, France}
\author{Alexander Farutin}
\affiliation{LIPhy, CNRS, F-38000, Grenoble, France}
\affiliation{LIPhy, Univ. Grenoble Alpes, F-38000, Grenoble, France}
\affiliation{Experimental Physics, Saarland University, 66123 Saarbrücken, Germany}
\author{Chaouqi Misbah}
\affiliation{LIPhy, CNRS, F-38000, Grenoble, France}
\affiliation{LIPhy, Univ. Grenoble Alpes, F-38000, Grenoble, France}
\author{Lionel Bureau}\email{lionel.bureau@ujf-grenoble.fr}
\affiliation{LIPhy, CNRS, F-38000, Grenoble, France}
\affiliation{LIPhy, Univ. Grenoble Alpes, F-38000, Grenoble, France}

\begin{abstract}%
The unique ability of a red blood cell to flow through extremely small microcapillaries depends on the viscoelastic properties of its membrane. Here, we study {\it in vitro} the response time upon flow startup exhibited by red blood cells confined into microchannels. We show that the characteristic transient time depends on the imposed flow strength, and that such a dependence gives access to both the effective viscosity and the elastic modulus controlling the temporal response of red cells. A simple theoretical analysis of our experimental data, validated by numerical simulations, further allows us to compute an estimate for the two-dimensional membrane viscosity of red blood cells, $\eta_{mem}^{2D}\sim 10^{-7}$ N$\cdot$s$\cdot$m$^{-1}$. By comparing our results with those from previous studies, we discuss and clarify the origin of the discrepancies found in the literature regarding the determination of $\eta_{mem}^{2D}$, and reconcile seemingly conflicting conclusions from previous works.
\end{abstract}

\maketitle

\section{Introduction}
\label{sec:intro}

The study of blood flow properties is a highly active field of research, both experimentally \cite{Baskurt2} and theoretically \cite{Gompper1,Gompper2,Misbah1}. Identifying the relevant (bio)physical parameters controlling such properties is of great interest in the context of diagnosis of blood disorders, and also represents an important fundamental challenge in numerical and theoretical modeling of complex fluids. Blood is a suspension of cellular elements (Red and White Blood Cells, Platelets) in a carrier fluid (Plasma). It is known to behave as a non-Newtonian fluid, the complex rheology of which is mainly due to the presence of Red Blood Cells (RBC), its major cellular component \cite{Baskurt2}. The flow properties of blood are thus essentially governed by the concentration of RBCs (the so-called hematocrit), their mutual interactions, and their individual mechanical properties. Over the past fifty years, the latter have been the focus of numerous studies which aimed at characterizing or modeling the viscoelastic behavior of individual RBCs, and in particular that of the cell membrane which separates the outer suspending plasma from the inner cytoplasm \cite{Chien,Hochmuth,Kim,Betz,Fedosov,Suresh1,Suresh2,Dimi}. The membrane region comprises a lipid bilayer, an external glycocalyx layer, and an inner two-dimensional (2D) cytoskeleton composed of a spectrin network. The inner and outer layers are connected to the bilayer through transmembrane proteins. Mechanical properties of the membrane are typically described in terms of bending and expansion moduli of the bilayer, a 2D elastic shear modulus of the spectrin network ($\mu_s$), and a 2D membrane viscosity ($\eta_{mem}^{2D}$) \cite{Hochmuth,Kim}.

Aside from their influence on bulk rheology, the membrane properties  of RBCs are also of utmost importance in microconfined flows, i.e. in the microvascular network, where they control the ability of cells to adapt their shape and flow through channels having dimensions on the order of or smaller than the size of unstrained RBCs. Such a capacity of RBCs to flow through very small channels is central in their physiological function of oxygen delivery to tissue, and alterations of the viscoelasticity of RBCs are recognized to be associated with various pathologies \cite{Mohandas}. In this context, {\it in vitro} studies of microconfined flows are currently viewed as a potential diagnosis tool to discriminate between healthy and pathological cells \cite{Kwan,Guido1,Guido2}. 

Along this line, several works have investigated the steady-state behavior of confined RBCs, from the theoretical \cite{Secomb2,Secomb3}, numerical \cite{Gompper3,Misbah2,Misbah4} and experimental \cite{Guido2,Stone,Rosen,Bit,Hou} points of view, in order to probe {\it e.g.} the influence of mechanical properties on cell shape or velocity. 
On the other hand, a limited number of studies have investigated the behavior of RBCs in transient situations. Early studies have investigated the relaxation of RBC deformation following the removal of an applied mechanical stress: the pioneer work by Hochmuth {\it et al.} focused on shape recovery in micropipette experiments \cite{Evans1};  Baskurt and Meiselman later performed rheo-optical strain relaxation experiments upon cessation of shear of concentrated RBC suspension \cite{Baskurt}, and Bronkhorst {\it et al.} pioneered the use of optical tweezers to conduct shape recovery experiments on single cells \cite{Bronkhorst}. More recently, Guido and Tomaiuolo have performed measurements of RBC response time upon flow start in microchannels \cite{Guido1}, and Braunm{\"u}ller {\it et al.} have investigated RBC shape relaxation using microfluidic tools \cite{Franke}.

The above studies have provided consistent results for the timescale controlling shape relaxation (or establishment) of healthy RBCs: $\tau_{c}\simeq0.1-0.2$ s. Moreover, chemical or physical treatments known to affect the mechanical properties of RBCs were observed to clearly modify this characteristic time \cite{Baskurt,Guido1}. From such measurements, authors concluded that $\tau_c$ was indeed governed by the viscoelastic properties of RBCs. Following Evans and Hochmuth \cite{Evans2}, $\tau_c$ has been commonly related to the  mechanical properties of the cell as: $\tau_{c}=\eta_{mem}^{2D}/\mu_s$. Using typical values $\mu_s \sim 5-10 \, \mu$N$\cdot$m$^{-1}$ taken from the literature \cite{Evans1,Guido1}, a 2D membrane viscosity $\eta_{mem}^{2D}\sim 0.5-1\times 10^{-6}$ N$\cdot$s$\cdot$m$^{-1}$ has been computed from the measured $\tau_c$ \cite{Guido1,Guido2,Evans1,Baskurt}. Puzzlingly, such studies of the transient response of RBCs conclude to a value of the membrane viscosity that contrasts with that coming from other groups of experiments, which rather yield $\eta_{mem}^{2D} \lesssim 10^{-7}$ N$\cdot$s$\cdot$m$^{-1}$ \cite{Chien2,Tran,Rasia}.

The origin of such a discrepancy between the few experimental determinations of the membrane viscosity is a long-standing issue and is still an open question. However, the membrane viscosity is often a required input parameter, either for experimental data analysis \cite{Fischer} or in advanced numerical models of RBCs \cite{Secomb1,Suresh2,Gompper1}. In the perspective of, {\it e.g.}, quantitative numerical studies of the flow behaviour of suspensions of RBCs, there is therefore a clear need for a more accurate knowledge of the membrane viscosity.

We address this question in the present paper. We present an analytical model that describes the shape evolution of a RBC in shear flow, which we validate using 3D numerical simulations. We use this framework to analyze the results of startup flow experiments. The latter are performed in the spirit of the study by Tomaiuolo and Guido \cite{Guido1}: we extend the work of these authors, and probe the effect on $\tau_c$ of (i) the flow strength, (ii) the viscosity of the suspending fluid ($\eta_{out}$), (iii) a chemical treatment known to affect the intrinsic mechanical properties of RBCs. 

We find that $\tau_c$ depends on the flow strength, and exploit this dependence to extract, in an original way, {\it both} the effective viscosity and the elastic modulus which govern the characteristic transient time of RBCs. We then demonstrate that such an effective viscosity is not identical to the membrane viscosity, but can be used to determine the actual  $\eta_{mem}^{2D}$. Doing so, we obtain a value of $\eta_{mem}^{2D}$ which is in good agreement with the low values reported by Tran-Son-Tay {\it et al.} \cite{Tran}. By combining theoretical, numerical, and experimental efforts in this study of viscoelastic transients of RBCs, we are thus able to reconcile the seemingly conflicting results regarding membrane viscosity of RBCs.

\section{Theoretical framework}
\label{sec:theory}

\subsection{Qualitative discussion}
\label{subsec:theory}

The membrane of a RBC as well as the internal and external fluids are characterized by their respective viscosities $\eta_{mem}$,  $\eta_{in}$ and $\eta_{out}$, with $\eta_{mem}=\eta_{mem}^{2D}R$, where $R$ designates the radius of a sphere having the same surface area as that of a RBC. RBC deformation due to external flow is accompanied by dissipation in the three fluids. Since one expects the slowest mechanism to govern the dynamics, this entails that the three regions act as dashpots in parallel, so that the total effective dissipation coefficient can be written as $\eta_{eff}= \alpha \eta_{out} +\beta' \eta_{in}+\gamma' \eta_{mem}$ where $\alpha$, $\beta'$ and $\gamma'$ are dimensionless numbers to be specified below. Shape deformation  of a RBC occurs on a time scale (to be determined) of order $\tau_c$ and deformation is typically on the order of the RBC radius, $R$, so that the viscous tension created by RBC deformation is of order $\eta _{eff} R/\tau_c$ (this is a force per unit length). This force counterbalances the combined effects of the force due to external flow, of order $\eta_{out} V$ ($V$ being a typical applied velocity), and of the non-dissipative part due to elasticity of the cell. The membrane is endowed with bending and shear elastic  energy due to cytoskeleton. Calling $\kappa$ the bending modulus (having the dimension of an energy), the typical force per unit length associated with it is of order $\kappa/R^2$. The shear elastic modulus is denoted as $\mu_s$ and represents a force per unit length (a 2D shear elastic modulus), so that total elastic force per unit length is of order $K_{eff}\simeq \kappa/R^2 +\mu_s$. The force balance yields
\begin{equation}
\label{tau}
\tau_c^{-1}=\frac{K_{eff}+ \eta_{out}V}{\eta_{eff}R},
\end{equation}
Eq. \ref{tau} provides us with the typical time scale of deformation of a RBC.
Let us rewrite (\ref{tau}) as 
\begin{equation}\label{taup}\tau_c^{-1}=a_1V+a_0,\end{equation}
where 
\begin{eqnarray} 
&&a{_1}^{-1}= R \alpha ({1+\beta \lambda+\gamma\lambda'})\label{eq:a1}\\
&&a_{0}^{-1} =\frac{R\eta_{out}\alpha (1+\beta \lambda+\gamma\lambda')}{K_{eff}}\label{eq:a0}
\end{eqnarray}
with $\lambda=\eta_{in}/\eta_{out}$, $\lambda'=\eta_{mem}/\eta_{out}$, and where $\beta=\beta'/\alpha$ and $\gamma=\gamma'/\alpha$. 

The different coefficients above ($\alpha$, $\beta$, $\gamma$) can only be determined numerically (see below). However, when the shape of a cell is not very far from a sphere, an analytical calculation is possible \cite{Misbah2006,Danker2007,Vlahovska2007,Lebedev2008,Vlahovska2011,Danker2009}. 
This has been done for the case of vesicles in a linear shear flow \cite{Lebedev2008}, taking into account $\eta_{out}$, $\eta_{in}$ and $\eta_{mem}$, and yields $\alpha=16/3$, $\beta=23/32$, and $\gamma=1/2$. A similar study has been performed for a Poiseuille flow \cite{Danker2009}, incorporating only the contributions of the outer and inner fluid viscosities, which provides $\alpha\simeq 9$ and $\beta=76/85$ for the flow of a vesicle in a channel of internal radius 5 $\mu$m, as it is the case in our experiments.
We discuss in the following section our choice for these parameters.

\subsection{Numerical simulations}
\label{subsec:num}

We have performed a systematic numerical study of a model of RBC.
The imposed flow is given by
\begin{equation}
\label{Poiseuille}
\boldsymbol{V}^\infty(\boldsymbol{R})=\boldsymbol{e}_xV\left[1-\frac{y^2+z^2}{(D/2)^2}\right],
\end{equation} 
where $V$ is the velocity at the center of the flow, the $x$ axis is along the flow direction, while the $y$ and $z$ axes are along the transverse directions. We consider here an unbounded flow. The curvature of the imposed Poiseuille flow alone, without the no-slip boundary conditions at the channel walls, has proven to account for several experimental facts provided that the RBC is not too confined, as shown in \cite{Coupier2012}. $D/2$ in Eq. \ref{Poiseuille} specifies at which distance from the center the imposed velocity falls to zero.  $D/2$ was set equal to 5 $\mu$m as in the experiments described below. 

Numerical simulations for 3D model of RBC  are based on the boundary integral method, as originally described for vesicles in \cite{Biben2011}. In a recent work \cite{Farutin2014}, we have extended this study to include membrane shear elasticity mimicking the spectrin network of RBCs (membrane viscosity is not accounted for at present).  We have also made several new numerical improvements that allowed us to study very deflated shapes, as required for simulation of real RBCs.  We only provide here the main results concerning numerical determination of shapes and relaxation time scales, while details of the numerical techniques can be found in \cite{Farutin2014}. The RBC model takes into account both bending and shear elasticity. The Helfrich model is adopted for the bending energy, while the shear elasticity of the cytoskeleton is described using a  FENEM-like (Finite Extensibility Non-linear Elasticity) strain hardening model, which has revealed to capture some realistic features of RBCs (for details see \cite{Farutin2014}).

 The RBC was modeled as an inextensible membrane, endowed with a shear modulus $\mu_s=1.9\, \mu$N$\cdot$m$^{-1}$ and a bending modulus $\kappa=2.7\times 10^{-19}$ J. The stress-free state of the membrane was chosen as a biconcave shape with a surface area of $122\, \mu$m$^2$ and a volume of $82\, \mu$m$^3$ ({\it i.e.} $R\simeq 3\,\mu$m). Most simulations were performed using 5120 triangular elements to discretize the membrane (2562 vertices). The viscosity of the inner solution of the cell was chosen as $\eta_{in}=6$ mPa$\cdot$s \cite{Betz}. The viscosity of the suspending medium, $\eta_{out}$, and the flow rate were varied as in the experiments. The elastic and geometrical parameters of the cell were chosen in order to get a model consistent with optical tweezers experiments \cite{Mills2004}. Several additional simulations were performed with different values of $\mu_s$ and $\kappa$ in order to mimic the chemical treatment of the cell.
 
 The characteristic time of the cell was determined by monitoring the transient behaviour of the cell velocity upon flow startup: in a manner similar to what has been used for experimental data analysis, we define the characteristic time $\tau_c$ as the time needed for the cell to reach 99\% of its steady-state velocity.

\section{Materials and Methods}
\label{sec:methods}

\subsection{Materials}
\label{subsec:materials}

Fused silica capillaries of inner diameter 10 $\mu$m and outer diameter 150 $\mu$m were purchased from BGB Analytik (Germany).
Polydimethylsiloxane (PDMS, Dow Corning Sylgard 184) was obtained from Neyco (France). Bovine Serum Albumin (BSA), Phosphate Buffered Saline (PBS) tablets, dextran of average molecular weight 40 kDa, and diamide were purchased from Sigma Aldrich (France). Solutions of PBS 0.01M and pH 7.4 were prepared using 18.2 M$\Omega\cdot$cm ultrapure water. 

The viscosity of PBS and PBS+dextran solutions was measured on a cone/plate rheometer (Anton Paar MCR301), and found to be $\eta_{out}=1.5\pm0.15$ mPa$\cdot$s and 5$\pm0.5$ mPa$\cdot$s respectively for pure PBS and PBS + 10\%w of dextran 40kDa.

\subsection{Preparation of blood samples}
\label{subsec:prepa}

Blood samples from healthy donors were obtained through the \'Etablissement Fran\c{c}ais du Sang (Grenoble) and stored at 4$^{\circ}$C until use. Red blood cells were extracted from whole blood by successive washes in PBS solution and centrifugation. After each centrifugation, the supernatant was pipetted out and PBS was added to refill the tube. The washing/centrifugation cycle was repeated three times.

RBCs were used as such, or after being exposed to diamide: using a protocol akin to that described in \cite{Stone}, erythrocytes were incubated at room temperature for 30 minutes in a PBS solution containing 5mM of diamide, then washed in PBS before use in the flow cell.

\subsection{Experimental Setup}
\label{subsec:setup}

\begin{figure}[h]
$$
\includegraphics[width=3.25in]{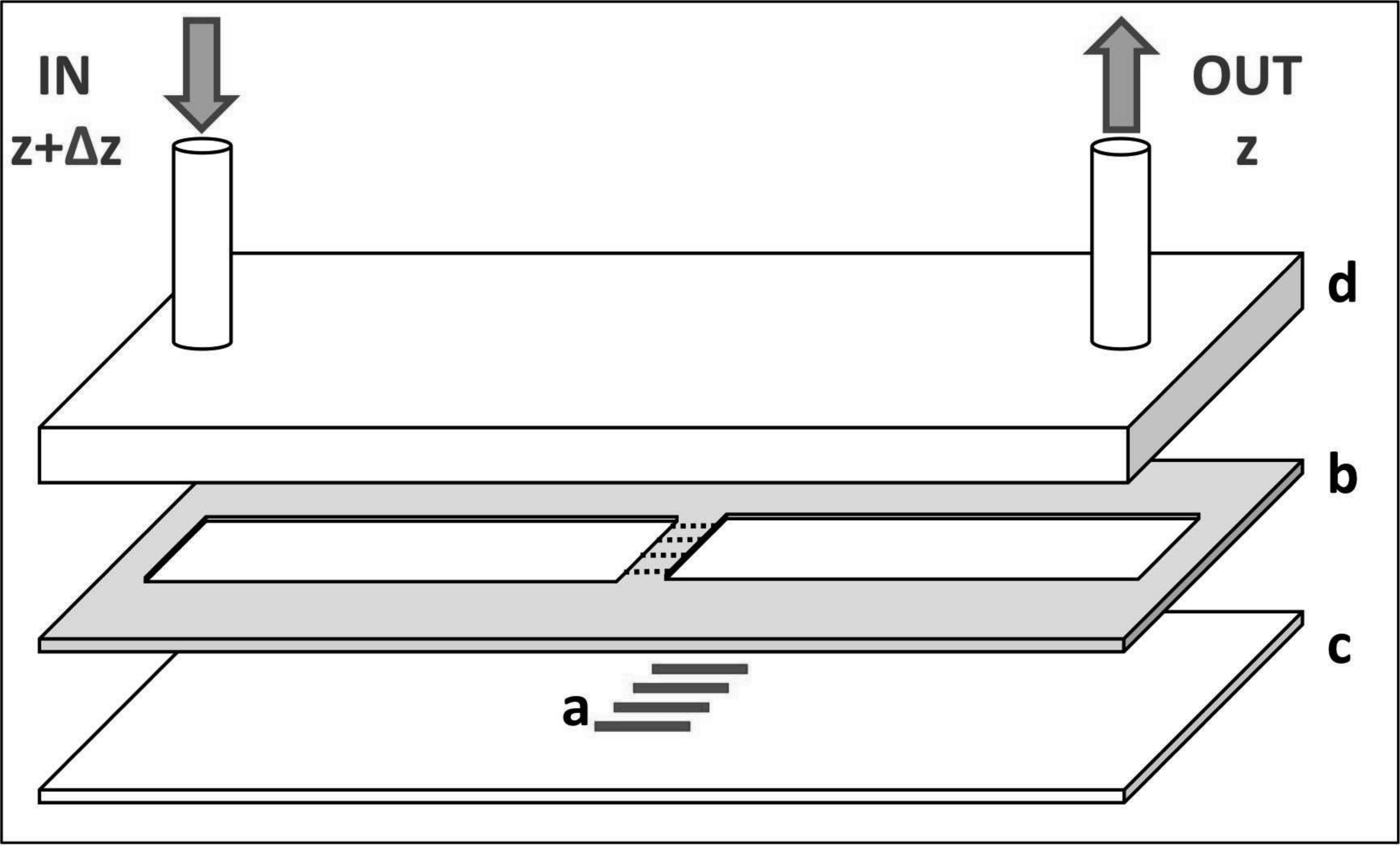}
$$
\caption{\label{fig:setup} Sketch of the flow cell used in the start-up experiments: (a) silica capillaries, (b) PDMS spacer with grooves, (c) glass coverslip, (d) polycarbonate plate.}
\end{figure}

Experiments were performed using a custom-built flow cell composed of the following elements, as illustrated in Fig. \ref{fig:setup}: 4 silica capillaries  were cut in 3-6 mm long segments and fitted into  4 grooves in the central band of a PDMS spacer. The spacer was then sandwiched between a glass coverslip and a clear polycarbonate plate, and the stack was clamped in an aluminum frame to ensure tight sealing of the cell. The inlet and outlet drilled into the polycarbonate upper plate were connected by silicone tubing to external reservoirs. A solenoid valve was connected between the inlet reservoir and the flow cell, in order to control the start-up of the flow, while the steady-state flow velocity was controlled by adjusting the height difference ($\Delta z$) between the liquid free surfaces in the inlet and outlet beakers. The response time upon startup of the whole setup (including valve, tubings and flow cell) was measured to be $\lesssim$ 10 ms (see Fig. \ref{fig:Vtransient}a below).

Before the experiments, the flow cell was filled with a PBS solution containing 0.2 \%w BSA, in order to coat the inner walls of the capillaries and prevent spurious adhesion of the cells to the walls of the microchannels. A RBC suspension was then added to the inlet reservoir and the concentration adjusted with PBS so that the initial bulk hematocrit was less than 1\%.

The flow cell was placed on the motorized stage of an inverted microscope (Olympus IX-71) equipped with a 100x oil-immersion objective. Image sequences of the flowing RBCs were acquired using a high speed camera (Phantom Miro4) at a frame rate of 4000 images per second.

\subsection{Experimental data analysis}
\label{subsec:expanal}

\begin{figure}[htbp]
$$
\includegraphics[width=3.25in]{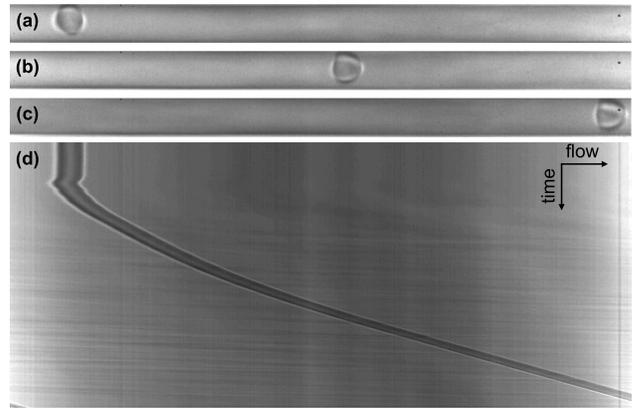}
$$
\caption{\label{fig:spacetime} Image of a RBC in a channel at rest (a), during transient (b), and at steady-state (c). (d) Space-time diagram built from the full image sequence: the diagram shows the averaged intensity along the flow direction (horizontally, flow is from left to right), as a function of time (vertically, from top to bottom). Image size is 167 $\mu$m $\times$ 85 ms. The steady state velocity is 3 mm$\cdot$s$^{-1}$, for a pressure drop of 2 kPa.}
\end{figure}

The recorded image sequences were analyzed using ImageJ software and Matlab toolboxes.
Image stacks showing the position of a RBC inside the capillary as a function of time were opened with ImageJ and cropped in order to keep only a rectangular region of interest (ROI), of height 10 $\mu$m and length 167 $\mu$m, enclosing the lumen of the capillary focused at its midplane (see Fig. \ref{fig:spacetime}a-c). The intensity of each image in the stack was averaged over the height of the ROI, and the result of this averaging was plotted as a function of time, in order to obtain a space-time diagram (see Fig. \ref{fig:spacetime}d). Such a space-time plot was further processed with a gradient filter and finally thresholded, in order to obtain a binary image showing the position of the front and rear parts of the flowing red blood cell as a function of time. This binary image was then processed with standard Matlab tools in order to extract the position of the front or rear ends as a function of time, $x(t)$, from which the instantaneous velocity $V_{RBC}(t)$ was computed. We have checked that no significant difference was obtained on $V_{RBC}(t)$ when computed from the position of the front or rear end, and that the obtained results were quantitatively in agreement with velocity measurements performed using Particle Imaging Velocimetry (PIV) tools from ImageJ. The method described above has the benefit of being much less time consuming than PIV.

The characteristic transient time ($\tau_c$) of RBCs upon flow startup was determined in two different ways: 

(i) we have estimated $\tau_c$ from visual inspection of the time-dependent shape of RBCs, defining $\tau_c$ as the time above which no change in RBC shape was visible by eye; 

(ii) we have computed the transient time from RBC velocity, defining $\tau_c$ as the time required for a RBC to reach 90\% of its steady-state velocity, as illustrated in Fig. \ref{fig:Vtransient}a below.

\section{Results}
\label{sec:results}

\subsection{Measurements of transient time}
\label{subsec:transientmeasure}

\begin{figure}[htbp]
$$
\includegraphics[width=3.25in]{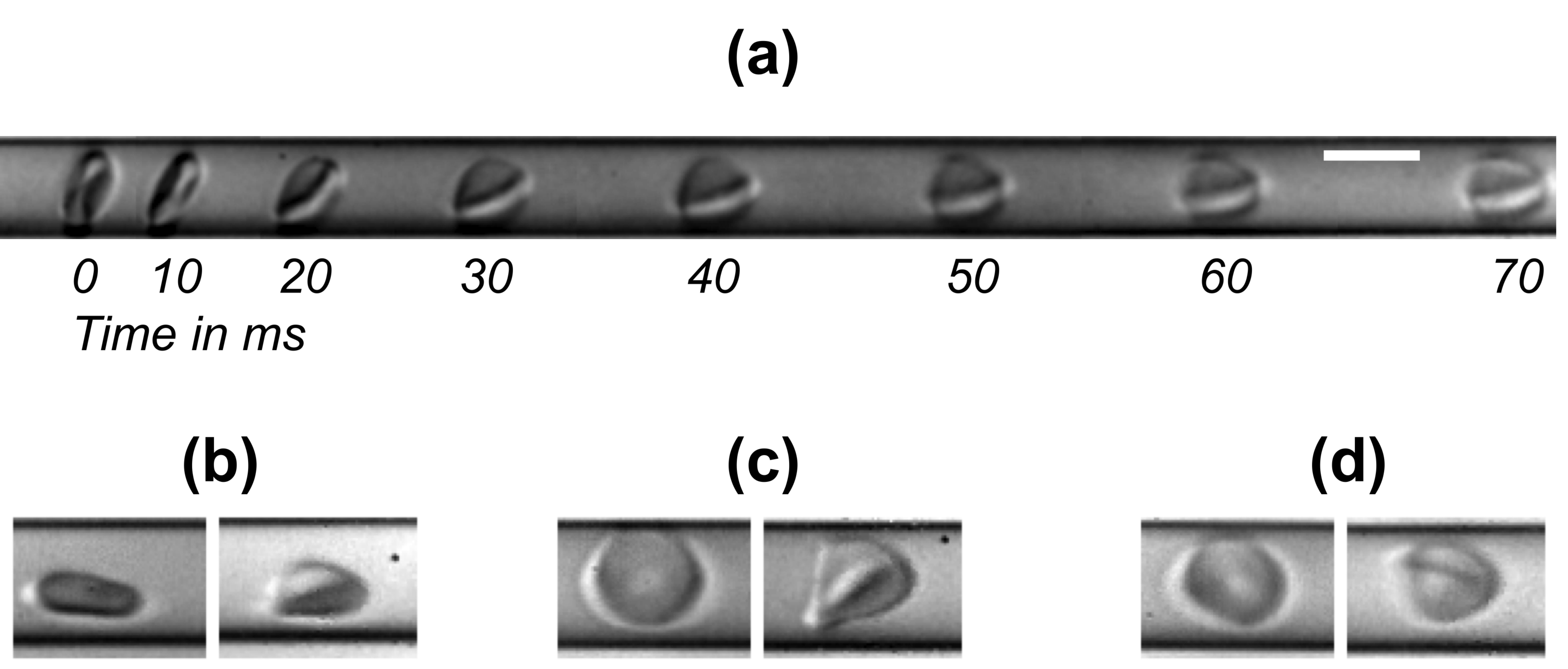}
$$
\caption{\label{fig:shapes} (a) Montage showing the time evolution of the shape of a RBC upon flow startup. Scale bar is 10 $\mu$m. Time is stamped below each image. The steady-state velocity reached by the RBC is $V_{max}=2.8$ mm$\cdot$s$^{-1}$. (b)-(d) illustration of the variety of initial RBC configurations (left image) and final steady-state shapes (right image) observed in our experiments. The three RBCs shown on (b)-(d) reach the same $V_{max}=0.88$ mm$\cdot$s$^{-1}$.}
\end{figure}

An example of cell behavior during flow startup is given in  Fig. \ref{fig:shapes}a. It can be seen that a RBC initially at rest in the channel gradually deforms with time, reaching a steady-state shape after a few tens of milliseconds. In the range of pressure drop explored in our work, the steady-state shape of the deformed RBC could be either parachute or slipper-like \cite{Skalak1969,Secomb1982,Misbah3}, as illustrated in Figs.  \ref{fig:shapes}b-d.

\begin{figure}[htbp]
$$
\includegraphics[width=3.25in]{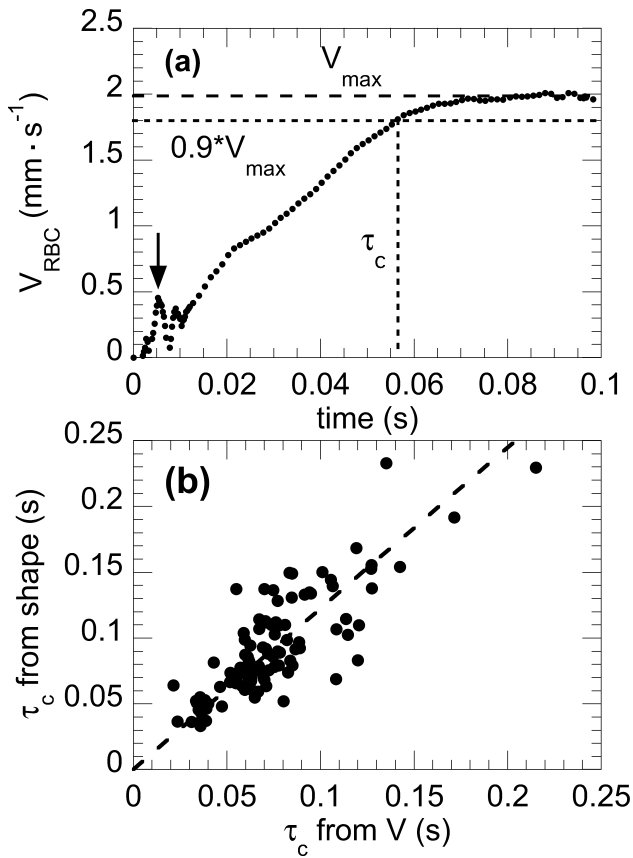}
$$
\caption{\label{fig:Vtransient} (a) Main panel: Time evolution of the velocity of a RBC going from rest at $t=0$ to a steady-state velocity $V_{max}=2$ mm$\cdot$s$^{-1}$. The arrow indicates the typical velocity overshoot observed upon opening of the solenoid valve, which lasts $\lesssim 10$ ms. Inset: dependence of $V_{max}$ on the imposed $\Delta P$. (b) $\tau_c$ determined from shape evolution as a function of $\tau_c$ computed from velocity transient. The dashed line is a linear ($y=ax$) fit to the data, with slope $a=1.22$.}
\end{figure}

The time dependence of the velocity of a RBC ($V_{RBC}$) is illustrated in Fig. \ref{fig:Vtransient}a. From such $V_{RBC}(t)$ curves we compute:

(i) the steady-state velocity of the cells ($V_{max}$), which depends, as expected for a Poiseuille flow, linearly on the imposed pressure drop $\Delta P$, as shown in the inset of Fig. \ref{fig:Vtransient}a, and 

(ii) the transient time $\tau_c$ as defined in Fig. \ref{fig:Vtransient}a.

In Fig. \ref{fig:Vtransient}b, we compare the characteristic time $\tau_c$ determined visually (from shape evolution) and from velocity transients, for all the data collected on RBCs suspended in PBS solution (the set of data corresponds to a total of about 100 cells). We find a good correlation between the times determined from the two methods, with $\tau_c$ computed from velocity being typically 20\% smaller. This difference arises, to a large extent, from the criterion we use to define $\tau_c$ from velocity transients, which underestimates the time to actually reach steady-state by 10-20\%.

The results presented in the rest of the paper correspond to $\tau_c$ determined from velocity transients. Data are presented as average values for $\tau_c$ and $V_{max}$, with error bars corresponding to $\pm$ one standard deviation, computed on groups of 8 to  12 RBCs.

\subsection{Dependence of $\tau_c$ on flow strength, external viscosity and diamide treatment}
\label{subsec:tauvsVandlambda}

\begin{figure}[htbp]
$$
\includegraphics[width=3.25in]{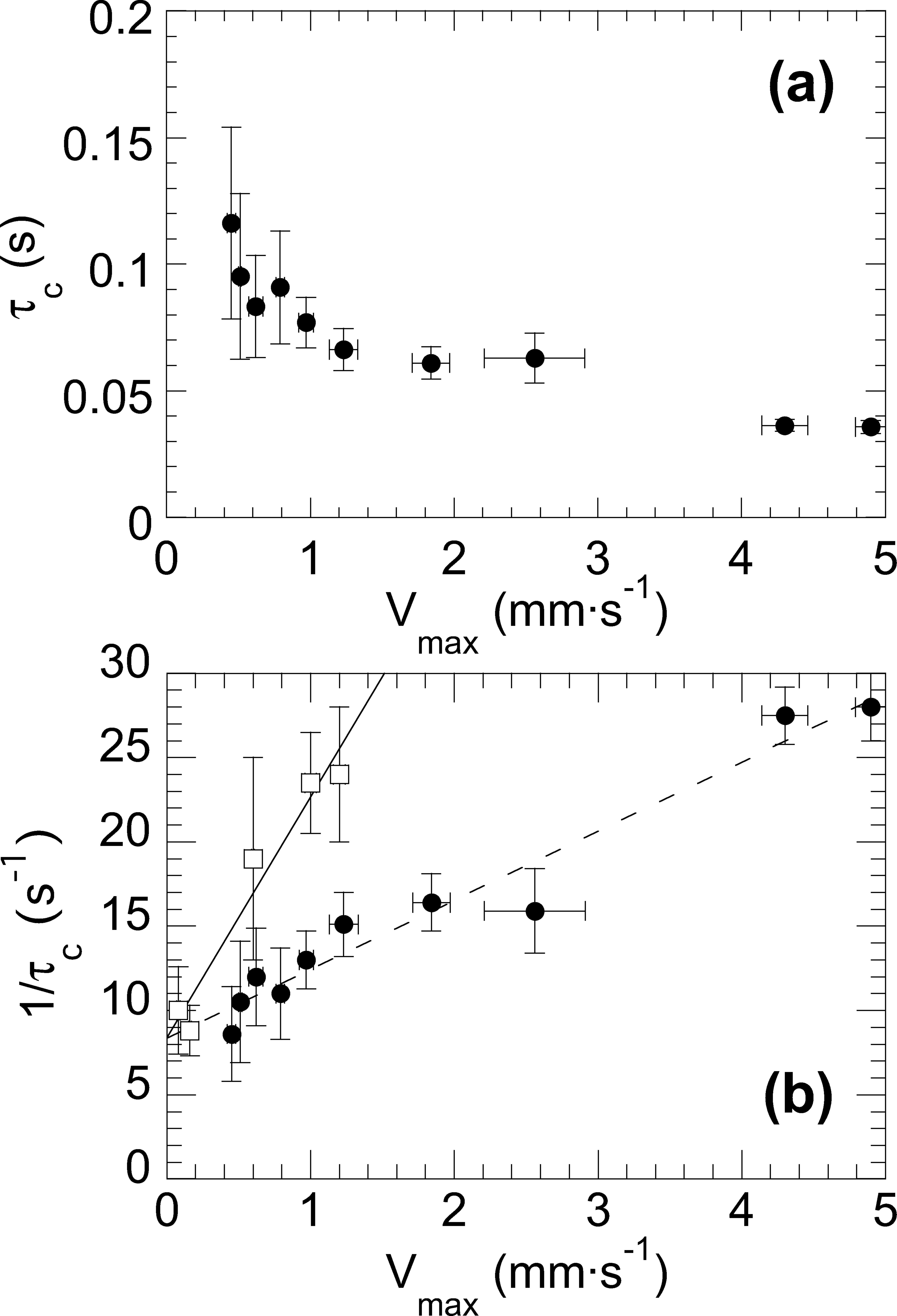}
$$
\caption{\label{fig:1surTau1}  Dependence of $\tau_c$ on experimental parameters. (a) $\tau_c$ as a function of $V_{max}$ for healthy RBCs suspended in PBS. (b) 1/$\tau_c$ as a function of $V_{max}$ for RBC in PBS ($\bullet$) and in PBS containing 10\%w of dextran 40kDa ({\tiny$\square$}). The dashed line is a linear fit, $\tau_c^{-1}=a_0+a_1V_{max}$, to the data obtained in PBS, with $a_0=8.3\pm0.8$ s$^{-1}$ and $a_1= 4.1\pm0.3$ mm$^{-1}$. The full line is a linear fit to the data obtained in PBS+dextran, with $a_0=8.4\pm1.4$ s$^{-1}$ and $a_1= 14.0\pm1.9$ mm$^{-1}$.}
\end{figure}

We observe that the characteristic time $\tau_c$ decreases as the flow strength increases. This is illustrated in Fig. \ref{fig:1surTau1}a, for RBCs suspended in  PBS solution, where it is seen that $\tau_c$ decreases from $\sim 0.1$ to 0.03 s as $V_{max}$ increases from 0.5 to 5 mm$\cdot$s$^{-1}$.

Plotting the inverse time $1/\tau_c$ as a function of $V_{max}$, we find that $1/\tau_c$ exhibits a linear increase with the steady-state velocity (Fig. \ref{fig:1surTau1}b).

Upon increasing the viscosity of the suspending medium from that of PBS solution ($\eta_{out}=1.5$ mPa$\cdot$s) to that of the 10\%w dextran solution ($\eta_{out}=5$ mPa$\cdot$s), we observe that the dependence of $1/\tau_c$ on $V_{max}$ remains linear, with a greater slope than for RBCs suspended in PBS and a similar value of the intercept at zero velocity (Fig. \ref{fig:1surTau1}b).

\begin{figure}[htbp]
$$
\includegraphics[width=3.25in]{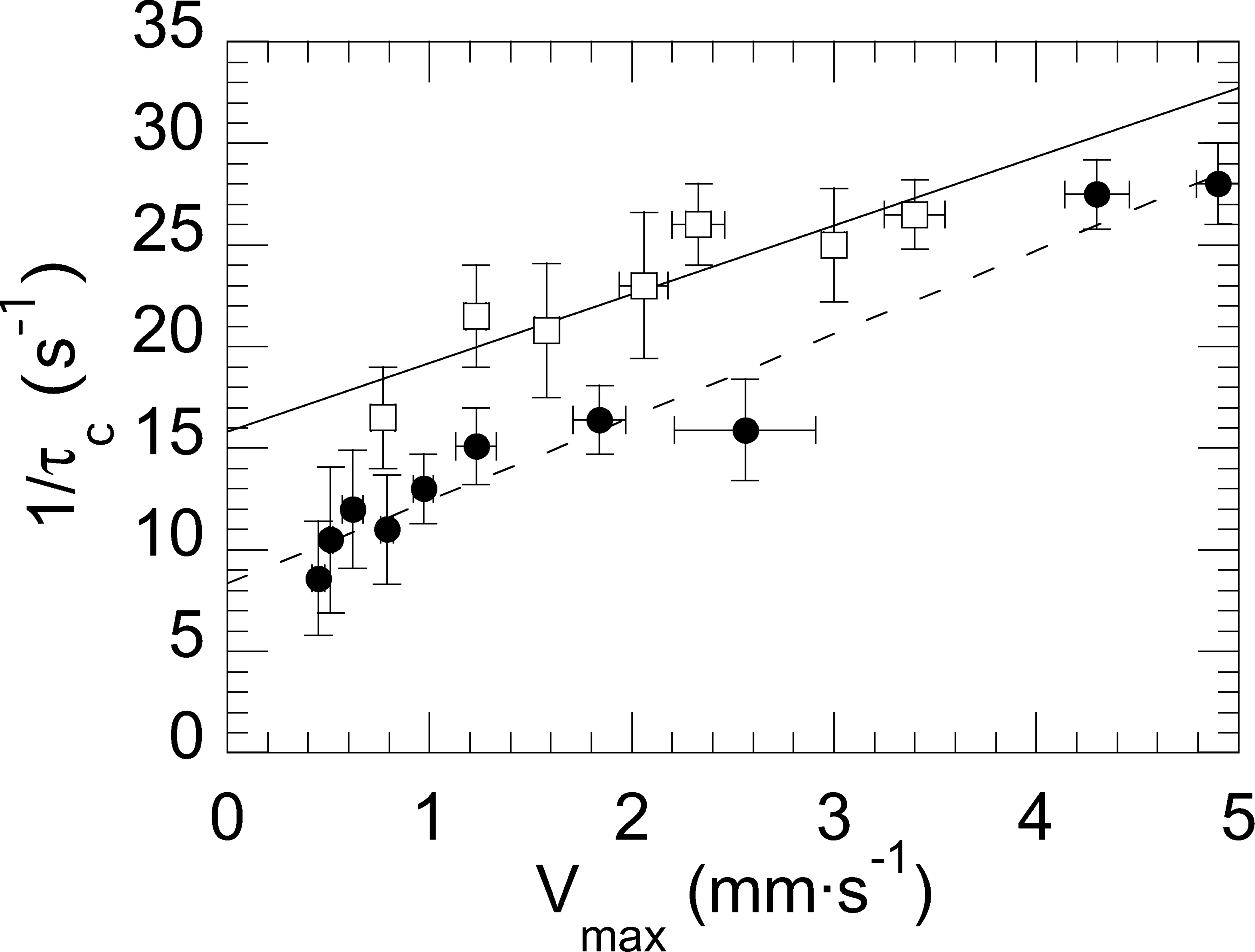}
$$
\caption{\label{fig:1surTau2} 1/$\tau_c$ as a function of $V_{max}$ for untreated RBCs suspended in PBS ($\bullet$), and diamide-treated RBCs in PBS ({\tiny$\square$}). Lines are best linear fits of the data for untreated (dashed line, with slope and intercept values as in Fig. \ref{fig:1surTau1}), and diamide-treated RBCs (full line, $a_0=15.8\pm1.6$ s$^{-1}$ and $a_1=3.4\pm0.7$ mm$^{-1}$).}
\end{figure}

Finally, we show in Fig. \ref{fig:1surTau2} the effect of diamide on transient time. It is seen that diamide-treated RBCs still exhibit an inverse transient time $1/\tau_c$ which increases linearly with $V_{max}$, with an overall upward shift of the curve with respect to that obtained for healthy RBCs. 

As discussed in detail below, all the above results are in good agreement with the simple heuristic argument given in section \ref{subsec:theory}. 

\subsection{Numerical results}
\label{subsec:NumRes}

\begin{figure}[htbp]
\begin{center}
\includegraphics[width=3.25in]{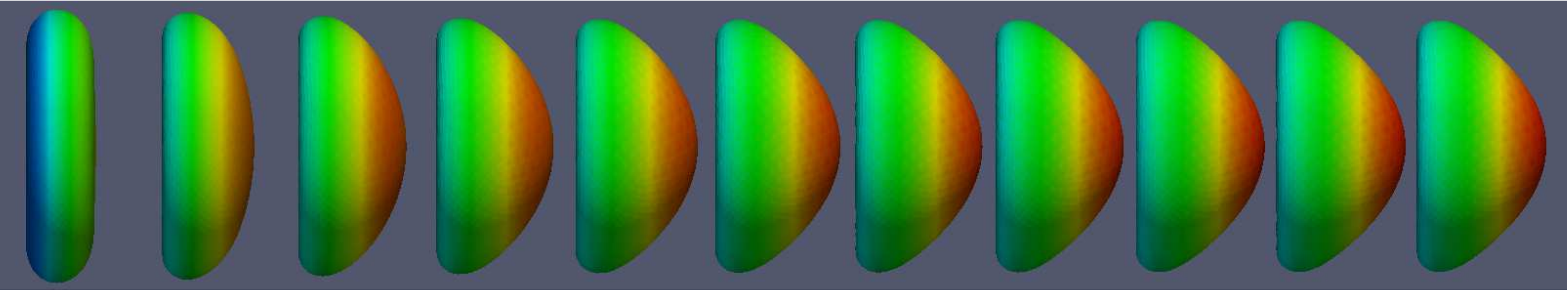}
\caption{\label{snapshot} Numerical results illustrating the shape transformation for a RBC in Poiseuille flow. Color by surface tension. $V=0.5$ mm$\cdot$s$^{-1}$, $\lambda=5.$}
\end{center}
\end{figure}

Fig. \ref{snapshot} shows a snapshot illustrating the temporal evolution of the shape of a RBC obtained from simulations, starting from a biconcave shape at rest.

\begin{figure}
\begin{center}
\includegraphics[width=3.25in]{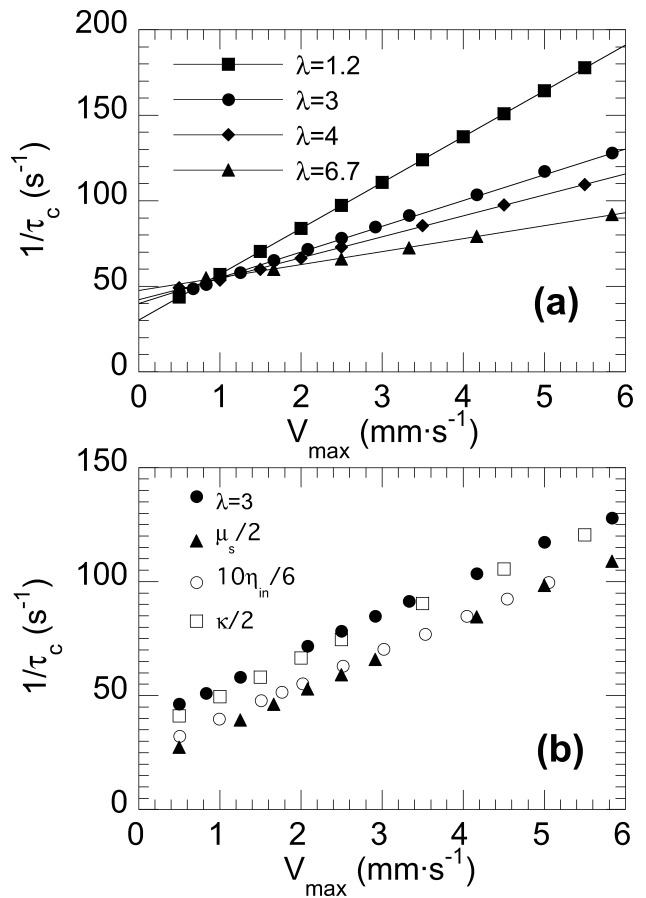}
\caption{\label{T} (a) Inverse relaxation time ($\tau_c^{-1}$) as a function of steady-state velocity ($V_{max}$), computed with $\eta_{in}=6$ mPa$\cdot$s, $\mu_s=1.9\, \mu$N$\cdot$m$^{-1}$ and $\kappa=2.7\times 10^{-19}$ J, for various viscosity contrasts: ({\tiny$\blacksquare$}) $\lambda=1.2$, ($\bullet$) $\lambda=3$, ({\small$\blacklozenge$}) $\lambda=4$, ($\blacktriangle$) $\lambda=6.7$. Full lines are best linear fits to the data ($\tau_c^{-1}=a_0^{num}+a_1^{num}V_{max}$).  
(b) $\tau_c^{-1}(V_{max})$ computed for $\lambda=3$ and parameters as in (a) ($\bullet$), with $\eta_{in}=10$ mPa$\cdot$s ($\circ$), with $\kappa=1.35\times10^{-19}$ J ({\tiny$\square$}); with $\mu_s=0.95\, \mu$N$\cdot$m$^{-1}$ ($\blacktriangle$).}
\end{center}
\end{figure}

The inverse relaxation time is plotted in Fig.\ref{T} as a function of the steady-state velocity. $1/\tau_c$ is seen to vary linearly with the flow rate, and to display a stronger dependence on $V_{max}$ for lower viscosity contrasts $\lambda$, {\it i.e.} for larger $\eta_{out}$ (see Fig. \ref{T}a). There is no apparent deviation from linearity neither for weak nor for strong flows in the range of explored velocities, in qualitative agreement with experiments.

As shown in Fig. \ref{T}b, for a given viscosity contrast $\lambda$, changing the value of the inner fluid viscosity, $\eta_{in}$, merely results in a vertical shift of the $\tau_c^{-1}(V_{max})$ curve, without affecting the slope. Data corresponding to an inner viscosity $\eta_{in}^2$ can be obtained from results computed for $\eta_{in}^1$ by simply rescaling $\tau_c^{-1}$ and $V_{max}$ by the ratio $\eta_{in}^1/\eta_{in}^2$.

The effect of varying the shear or the bending modulus is illustrated in Fig. \ref{T}b. Decreasing $\mu_s$ by a factor of two leaves the slope of the curve essentially unchanged, and results in a twofold decrease of the intercept at zero velocity. Decreasing $\kappa$ by a factor of two increases the slope by 10\% and lowers the offset at zero velocity by about 15\%.

\section{Discussion}
\label{sec:discussion}

\subsection{Numerical and analytical results}
\label{subsec:NumvsTh}

Fig. \ref{snapshot} illustrates the ability of the numerical model to reproduce the shape evolution of a RBC upon startup of a Poiseuille flow: the shape transition from discocyte to parachute obtained numerically is consistent with experimental observations such as that shown in Fig. \ref{fig:shapes}a. 

There is a strikingly good qualitative agreement between the simulations and the model presented in section \ref{sec:theory}: as inferred from a heuristic argument (Eq. \ref{tau}), a linear dependence of $1/\tau_c$ on $V_{max}$ is predicted numerically, the slope of this linear dependence is controlled by the effective  viscosity, hence by the ratio $\lambda$, while the elastic moduli affect only the value of $\tau_c^{-1}$ at $V_{max}=0$. The magnitude of the effects produced by varying either $\mu_s$ or $\kappa$ indicates that  the elastic behavior is essentially governed by $\mu_s$ (Fig. \ref{T}b).

\begin{figure}[htbp]
\begin{center}
\includegraphics[width=3.25in]{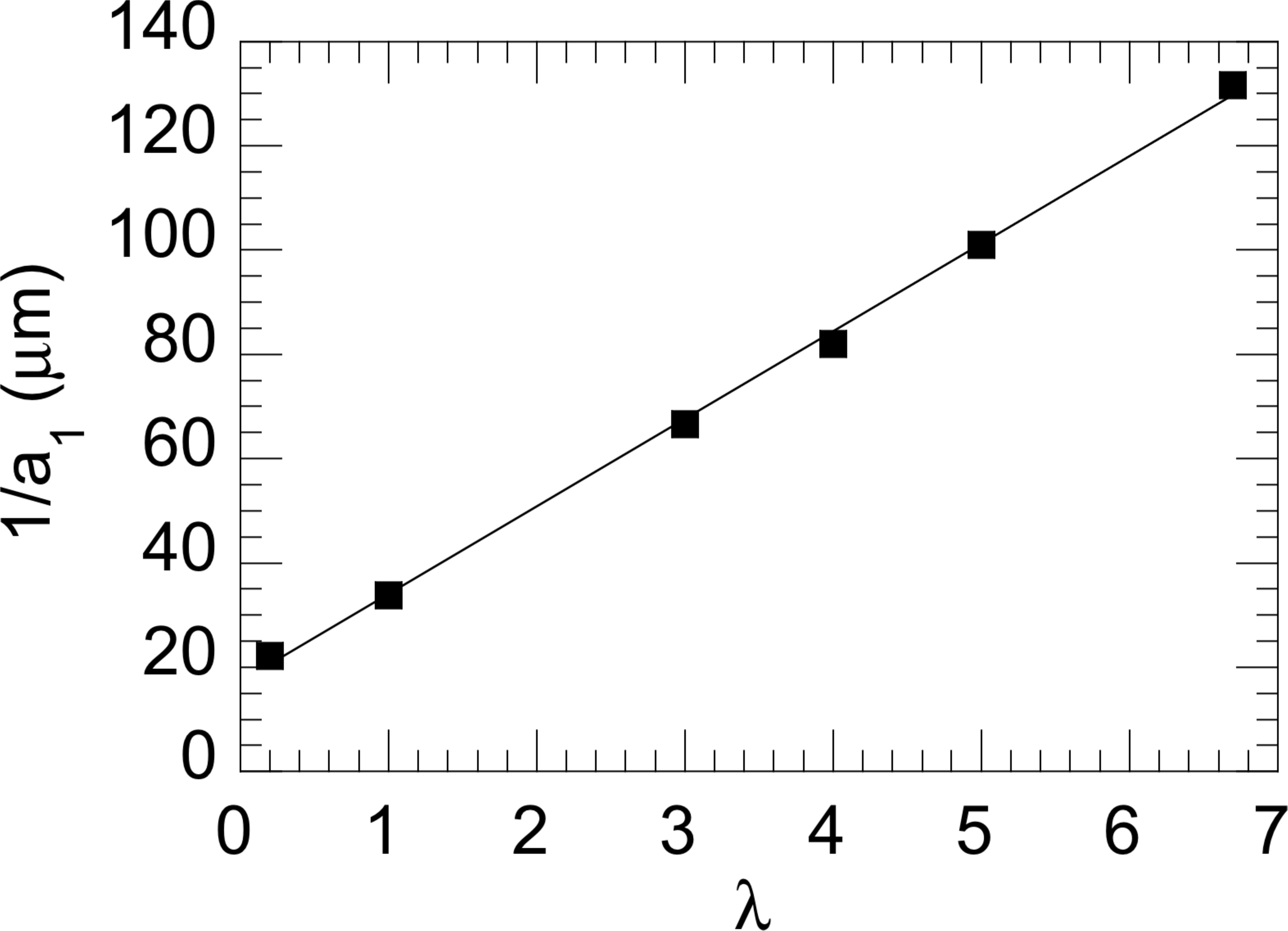}
\caption{\label{1sura1} Inverse slope, $1/a_1$, as a function of viscosity contrast $\lambda$ ({\tiny$\blacksquare$}). The full line is the best linear fit to the data, with a value of the intercept at $\lambda=0$ of 17.2 $\mu$m and a slope of 16.8 $\mu$m.}
\end{center}
\end{figure}

Such an agreement between numerics and theory prompts us to further exploit numerical simulations. Fig. \ref{T}a shows that the slope of $1/\tau_c$ $vs$ $V_{max}$, called $a_1$ in Eq. \ref{taup}, depends on the viscosity contrast $\lambda=\eta_{in}/\eta_{out}$. In Fig. \ref{1sura1}, we plot the inverse slope, $a_1^{-1}$, obtained from numerical simulations, as a function of $\lambda$. It is seen that $a_1^{-1}$ increases linearly with $\lambda$. This is again fully consistent with theoretical predictions (see Eq. \ref{eq:a1}). A linear fit of $a_1^{-1}(\lambda)$ therefore allows us to determine the parameters $\alpha$ and $\beta$, which set the relative contributions of $\eta_{out}$ and $\eta_{in}$ to the effective viscosity (see Eq. \ref{eq:a1}): we get $\alpha R=17.2\, \mu$m and $\alpha R \beta=16.8\, \mu$m, which yields, with $R=3\mu$m, $\alpha=5.7$ and $\beta=0.98$. These values are to be compared with those predicted by theories, namely  $\alpha\simeq 5.3$ and $\beta\simeq 0.72$ for the case of a linear shear flow \cite{Lebedev2008}, and $\alpha\simeq 9$ and $\beta\simeq 0.9$ for vesicles in a Poiseuille flow \cite{Danker2009}. The fair agreement between the parameters determined numerically and their theoretical values is all the more satisfactory that the analytical treatments proposed in \cite{Lebedev2008} and \cite{Danker2009} are based on spherical harmonics expansions that are strictly valid only for small deformations around a reference spherical shape, whereas numerical simulations handle the case of an initially biconcave, highly deflated, cell undergoing large deformations under flow. This provides strong support to the fact that analytical models such as those developed in \cite{Lebedev2008} and \cite{Danker2009}  are robust and able to capture the main contributions to dissipation during vesicle flow.

On this basis, we will now analyze our experimental data within the proposed theoretical framework. Since numerical simulations account for more realistic shapes and deformations of the cells than theoretical models, we will use in the following the values of the parameters $\alpha$ and $\beta$ determined from the numerical results. As far as the parameter $\gamma$ is concerned, it cannot be determined from simulations since membrane viscosity is not included in the numerical model. Still, the reasonable agreement between the numerical and theoretical values obtained for $\alpha$ and $\beta$ suggests that analytical parameters are of the correct order of magnitude, and we will therefore use the theoretical value $\gamma=1/2$ \cite{Lebedev2008}. A more accurate estimate of $\gamma$, {\it i.e.} of the exact weight of $\eta_{mem}$ in $\eta_{eff}$, would call for more advanced simulations accounting for membrane viscosity.

\subsection{Interpretation of experimental results for healthy RBC}
\label{subsec:ExpInter}

We first compare our results with those obtained by  Tomaiuolo and Guido \cite{Guido1}: most of the results presented in their study have been obtained at a single  steady-state velocity of 1 mm$\cdot$s$^{-1}$, at which $\tau_c=0.09$ s is found for healthy RBCs in PBS. At the same velocity, we measure a characteristic time $\tau_c=0.08\pm0.01$ s, which is in excellent quantitative agreement. 

Besides, as predicted from a simple heuristic argument (see Eqs. \ref{tau} and \ref{taup}), we find experimentally that $1/\tau_c$ is linear with $V_{max}$ (Fig. \ref{fig:1surTau1}).
Linear fits of the data obtained in PBS and PBS+dextran yield the values of the intercept at origin ($a_0$) and slope ($a_1$) reported in Table \ref{tab}. 
We observe roughly a threefold increase in the slope when $\eta_{out}$ is multiplied by 3. Recalling that $a_1=\eta_{out}/(\eta_{eff}R)$ (Eq. \ref{tau}), this suggests that $\eta_{eff}$ is only marginally affected by the external viscosity for $\eta_{out}$ in the range $1.5-5$ mPa$\cdot$s, hence that the contribution of $\eta_{out}$ to $\eta_{eff}$ is weak. This is also consistent with the fact that the change in $\eta_{out}$ does not modify significantly the value of $a_0=K_{eff}/(\eta_{eff}R)$. In our experiments, the main effect of changing $\eta_{out}$ is therefore to affect the magnitude of the hydrodynamic forces ($\eta_{out}V$) applied to the RBC. 

From $a_0$ and $a_1$, we compute the effective viscosity and elastic modulus which are given in Table \ref{tab}. We thus get a  value of $\eta_{eff}\simeq 120$ mPa$\cdot$s which is in good agreement with the one reported in a recent study by Betz {\it et al.}, namely $\eta_{eff}\simeq100$ mPa$\cdot$s \cite{Betz}. The value of $K_{eff}\simeq3\,\mu$N$\cdot$m$^{-1}$ is consistent with that of 2.5$\pm0.4 \,\mu$N$\cdot$m$^{-1}$ reported for $\mu_s$  by  Henon {\it et al.} \cite{Henon}. This suggests that, in the experimental conditions used in the present study, $K_{eff}$ is governed by the shear elastic modulus. This is in agreement with our numerical results, which show that the value of $\tau_c^{-1}$ at $V_{max}=0$ is essentially controlled by  $\mu_s$ (Fig. \ref{T}b).

Furthermore, from the values of $a_1$ and using Eq. \ref{eq:a1}, we make the following estimates for the membrane viscosity: $\eta_{mem}\simeq 20-28$ mPa$\cdot$s (see Table \ref{tab}). From such an estimate, we conclude that, for an outer fluid viscosity in the physiological range ($\eta_{out}\sim 1$ mPa$\cdot$s), $\eta_{in}$ and $\eta_{mem}$ are the two major contributions to the effective viscosity: $\alpha(\beta\eta_{in}+\gamma\eta_{mem})\simeq 100$ mPa$\cdot$s $\simeq \eta_{eff}$. Finally, if we compute a 2D membrane viscosity as $\eta_{mem}^{2D}=\eta_{mem}R$, we get $\eta_{mem}^{2D}\simeq 0.6-0.85\times 10^{-7}$ N$\cdot$s$\cdot$m$^{-1}$. 

To summarize, our results regarding the order of magnitude of $\tau_c$ are consistent with those from previous works on the transient response of RBCs \cite{Evans1,Guido1,Guido2}, but on the other hand we estimate a membrane viscosity $\eta_{mem}^{2D}$ which is about one order of magnitude lower than the values reported in these studies ({\it i.e.} $\eta_{mem}^{2D}\sim 10^{-6}$ N$\cdot$s$\cdot$m$^{-1}$ \cite{Evans1}). However, we note that the latter are based on an analysis proposed initially by Evans and Hochmuth \cite{Evans2}, who developed a model assuming that the only dissipation relevant to RBC shape recovery is associated to the membrane viscosity. They thereby neglected dissipation in the inner and outer fluids, and estimated $\eta_{mem}^{2D}\simeq \tau_c\mu_s$.  By contrast, here we make no such assumption. We find that (i) the effective viscosity is a combination of $\eta_{mem}$, $\eta_{in}$ and $\eta_{out}$,  and (ii) each viscosity entering $\eta_{eff}$ has a weight which is larger than 1 ($\alpha\simeq 6$). The latter point, which is an outpout of our theoretical and numerical analysis, is consistent with the {\it ad hoc} assumption made by Betz {\it et al.} in their modeling of membrane fluctuations \cite{Betz}. Taking into account all the viscous contributions and their respective weights in our data analysis, we get values for $\eta_{mem}^{2D}$ in very good agreement with those reported by Tran-Son-Tay {\it et al.} \cite{Tran}. These authors deduced $\eta_{mem}^{2D}$ from the tank-treading frequencies of RBCs under shear, using a model incorporating all dissipation sources, and obtained values which lie between 0.5 and 1.2$\times 10^{-7}$ N$\cdot$s$\cdot$m$^{-1}$. Our work, fully consistent with the one of Tran-Son-Tay {\it et al.} \cite{Tran}, thus strongly suggests that the assumption made by Evans and Hochmuth leads to an overestimate of $\eta_{mem}^{2D}$ and is at the origin of the apparent discrepancy between the reported values of RBC membrane viscosity. 

\begin{table}[htbp]
\begin{tabular}{cccccc}
$\eta_{out}$ & $a_0$ & $a_1$ & $\eta_{eff}$ & $K_{eff}$ & $ \eta_{mem}$ \\
(mPa$\cdot$s) & ($s^{-1}$) & (mm$^{-1}$) & (mPa$\cdot$s) & ($\mu$N$\cdot$m$^{-1}$) & (mPa$\cdot$s) \\ \hline
1.5 & $8.3\pm0.8$ & $4.1\pm0.3$ & $122\pm21$ & $3.0\pm0.8$ & $28\pm8$ \\ \hline
  5 & $8.4\pm1.4$ & $14.0\pm1.9$ & $119\pm28$ & $3.0\pm1.2$ & $20\pm11$
\end{tabular}
\caption{\label{tab}Values of $a_0$ and $a_1$ obtained from a linear fit of the experimental data for RBCs suspended in PBS ($\eta_{out}=1.5$ mPa$\cdot$s) and in PBS+dextran ($\eta_{out}=5$ mPa$\cdot$s). The corresponding values of $\eta_{eff}$ and $K_{eff}$ have been computed from Eqs. \ref{tau} and \ref{taup} with $R=3\,\mu$m. The membrane viscosity $\eta_{mem}$ has been calculated from $\eta_{eff}$ using Eq. \ref{eq:a1} with $R=3\,\mu$m, $\alpha=5.7$, $\beta=0.98$, $\gamma=0.5$ and $\eta_{in}=6$ mPa$\cdot$s.}
\end{table}

\subsection{Comparison between experiments and numerical simulations}
\label{subsec:numvsexp}

Although numerical simulations show a linear dependence of $1/\tau_c$ on $V_{max}$, in qualitative agreement with experiments, we observe that both the slope ($a_1^{num}$) and the extrapolated value of  $1/\tau_c$ at zero velocity ($a_0^{num}$) predicted numerically are larger than the experimental values ($a_1^{exp}$ and $a_0^{exp}$). This is shown in the summary table below (Table \ref{tab2}). 

Following the data analysis proposed in the previous section, the difference in slopes can straightforwardly be attributed to the contribution of membrane viscosity to the effective viscosity. Indeed, we have used the numerical results regarding $a_1^{num}(\lambda)$ to determine the parameters $\alpha$ and $\beta$ (Fig. \ref{1sura1}), and then fitted the experimental data using Eqs. \ref{taup} and \ref{eq:a1} in order to evaluate $\eta_{mem}$, so that the difference between the inverse slopes merely reads $1/a_1^{exp}-1/a_1^{num}=\alpha\gamma R\eta_{mem}/\eta_{out}$. Using the latter expression to compute $\eta_{mem}$ yields exactly the same values reported before for the membrane viscosity.

Now, a much more stringent test is to check whether the difference in the offsets ($a_0$) can be attributed to the contribution of membrane viscosity, because in contrast to $a_1^{num}$, $a_0^{num}$ was not used to determine parameters of the model. From Eq. \ref{eq:a0}, we expect:
\begin{equation}
\label{eq:diffa0}
\frac{1}{a_0^{exp}}-\frac{1}{a_0^{num}}=\frac{\alpha\gamma\eta_{mem}R}{K_{eff}} 
\end{equation}
In order to check the above equality, we estimate $K_{eff}$ from Eq. (\ref{eq:diffa0}) and compare it to the effective modulus determined in the previous section. Using the values of $a_0$ reported in Table \ref{tab2} and of $\eta_{mem}$ determined above, we get: $K_{eff}=2-2.5\,\mu$N$\cdot$m$^{-1}$ respectively for $\lambda=1.2$ ($\eta_{out}=5$ mPa$\cdot$s) and $\lambda=4$ ($\eta_{out}=1.5$ mPa$\cdot$s). This is in quantitative agreement with the value of $K_{eff}\simeq 3\pm1\, \mu$N$\cdot$m$^{-1}$  determined from experimental data, and therefore supports the fact that discrepancies between simulations and experiments are due to the missing contribution of membrane viscosity in the numerical model.

\begin{table}[h]
\begin{tabular}{ccccc}
$\lambda$ & $a_0^{num}$ ($s^{-1}$)& $a_1^{num}$ (mm$^{-1}$) & $a_0^{exp}$ ($s^{-1}$)& $a_1^{exp}$  (mm$^{-1}$)\\ \hline
  4  & 42.1 & 12.3 & 8.3 & 4.1\\ \hline
1.2   & 30.3 & 26.8 & 8.4 & 14.0
\end{tabular}
\caption{\label{tab2}  Comparison of numerical and experimental values of $a_0$ and $a_1$.}
\end{table}

\subsection{Effect of diamide treatment}
\label{subsec:expdiamide}

In Fig. \ref{fig:1surTau2}, we see that treating RBCs with diamide mainly results in an upward shift of the curve $\tau_c^{-1}(V_{max})$, {\it i.e.} diamide-treated RBCs exhibit a shorter transient time than healthy ones. From a linear fit to the data, we obtain $\eta_{eff}=147\pm45$ mPa$\cdot$s and $K_{eff}=7.0\pm2.8 \, \mu$N$\cdot$m$^{-1}$. The major effect of diamide  is therefore an increase of the effective elastic modulus $K_{eff}$, by roughly a factor of 2 with respect to healthy RBCs. Diamide is known to affect the spectrin network of RBCs, by creating crosslinks between proteins through the formation of disulphide bonds \cite{Fischer2}. Such a crosslinking process has been reported to induce a stiffening of RBCs \cite{Fischer2,Johnson,Rodrigues,Shin}, via an increase in the membrane shear modulus \cite{Fischer2}. Our results concerning the effect of diamide on RBC transient time are therefore fully consistent with these previous studies. Furthermore, attributing the increase of $K_{eff}$ to an increase in $\mu_s$ is consistent with the numerical results presented in section \ref{subsec:NumRes}. However, Forsyth {\it et al.}, from which we have adapted the protocol for diamide treatment in the present work, recently reported that, in microfluidics experiments relying on cell stretching during transient confinement  in a channel constriction, no difference in elongation could be detected between healthy and diamide-treated RBCs \cite{Stone}. We believe that the apparent contradiction between our finding and the conclusions of reference \cite{Stone} merely arises from the very different time scales probed in the two studies. Indeed, while we characterize the effect of diamide by measuring the transient viscoelastic time of RBCs, of order $50$ ms, Forsyth {\it et al.} probe the elongation of RBCs during a transient confinement that lasts between 1 and 5 ms only (calculated from the flow speeds and constriction length reported in \cite{Stone}). The duration of confinement in the work of Forsyth {\it et al.} is therefore much shorter than the time required for RBCs to reach their steady shape. This, added to the fact that we have used both a slightly larger diamide concentration and treatment time in our study, is likely to be at the origin of the differences between our work and the one reported in \cite{Stone}.

\section{Conclusions}

We have performed a study of the characteristic viscoelastic time $\tau_c$ of red blood cells. We have combined theory and numerical simulations in order to establish a framework for the analysis of  flow startup experiments on RBCs confined into microchannels. We have obtained experimental values of $\tau_c$  in quantitative agreement with those from previous studies, measured either from flow startup \cite{Guido1}, relaxation after cessation of shear \cite{Baskurt}, or micropipette experiments \cite{Evans1}. Moreover, we have shown that probing the dependence of $\tau_c$ on flow strength allows us to determine both the effective viscosity ($\eta_{eff}$) and elastic modulus ($K_{eff}$), and have obtained values for these two quantities which are consistent with those from other works \cite{Betz,Henon}. Most importantly, we have shown that, in contrast to the assumption made by Evans and Hochmuth \cite{Evans2} and commonly used later \cite{Guido1,Guido2,Evans1,Baskurt}, $\eta_{eff}$ is not equal to the membrane viscosity $\eta_{mem}$.  We have identified, from theory and simulations, the relative contributions of the membrane,  inner ($\eta_{in}$) and outer ($\eta_{out}$) fluid viscosities  to the overall effective viscosity, and  used this in order to compute the value of the membrane viscosity from our experimental data. Doing so, we obtain $\eta_{mem}$ in the range $20-28$ mPa$\cdot$s, which translates into a 2D viscosity $\eta_{mem}^{2D}=0.6-0.85\times 10^{-7}$ N$\cdot$s$\cdot$m$^{-1}$. We conclude that the difference of up to one order of magnitude that can be found in the literature regarding $\eta_{mem}$ only results from the fact that assuming $\eta_{eff}\equiv \eta_{mem}$ leads to overestimate  $\eta_{mem}$ by about a factor of ten. The present work thus reconciles previous contrasting reports about RBC membrane viscosity, and provides a range of values for $\eta_{mem}^{2D}$ which is in excellent quantitative agreement with the result of Tran-Son-Tay {\it et al.} \cite{Tran}. Moreover, we have demonstrated that measuring $\tau_c$ as a function of  flow strength provides valuable information not only on dissipation, but also on the elastic response of RBCs, and is sensitive to elasticity alterations such as those caused by diamide. This type of experimental investigations, analysed within the proposed framework, should therefore prove to be useful in discriminating how physical or biological factors may affect red blood cell elasticity or viscosity.

\section{Acknowledgments}

We acknowledge financial support from CNES (Centre National d'Etudes Spatiales), ESA (European Space Agency), and Universit\'e Franco-Allemande, Coll{\`e}ge doctoral ``Liquides Vivants'' (AF).


\bibliographystyle{bj}
\bibliography{startup}

\end{document}